\documentclass{tPHM2e}

\usepackage{epstopdf}
\usepackage{subfigure}

\theoremstyle{plain}

\theoremstyle{definition}

\theoremstyle{remark}

\newcommand{\beq}{\begin{equation}}
	\newcommand{\eeq}[1]{\label{#1}\end{equation}}
\newcommand{\B}[1]{{\bm{#1}}}
\newcommand{\brk}[1]{\left(#1\right)}

\newcommand{\matrixII}[4]{\left(\begin{array}{cc}#1&#2\\#3&#4\end{array}\right)}

\renewcommand{\exp}[1]{e^{#1}}

\newcommand{\xvec}{\mathbf{x}}
\newcommand{\Qvec}{\mathbf{Q}}

\newcommand{\Cof}[1]{\text{Cof}(#1)}

\usepackage{color}

\begin{document}


\articletype{Article}

\title{Microalloying and the mechanical properties of amorphous solids}

\author{
\name{H. George E. Hentschel\textsuperscript{a,b}
, Michael Moshe\textsuperscript{c,d}$^{\ast}$\thanks{$^\ast$Corresponding author. Email: mmoshe@syr.edu}, Itamar Procaccia\textsuperscript{a} and Konrad Samwer\textsuperscript{e} }
\affil{\textsuperscript{a}Weizmann Institute of Science,  Rehovot 76100, Israel ;
\textsuperscript{b}Department of Physics, Emory University, Atlanta, Georgia, USA; \textsuperscript{c}Department of Physics, Syracuse University,
Syracuse, New York 13244-1130, USA ; \textsuperscript{d}Department of Physics, Harvard University, Cambridge, MA 02138, USA ; \textsuperscript{e} Department of Physics, University of G\"ottingen, Germany}
}

\maketitle

\begin{abstract}
The mechanical properties of amorphous solids like metallic glasses can be dramatically
changed by adding small concentrations (as low as 0.1\%) of foreign elements. The glass-forming-ability, the ductility,
the yield stress and the elastic moduli can all be greatly effected. This paper presents theoretical considerations with
the aim of explaining the magnitude of these changes in light of the small concentrations involved. The theory is built
around the experimental evidence
that the microalloying elements organize around them a neighborhood that differs from both the crystalline and the glassy
phases of the material in the absence of the additional elements. These regions act as  {\em isotropic} defects that in
unstressed systems modify the shear moduli. When strained, these defects
interact with the incipient plastic responses which are quadrupolar in nature. It will be shown that this
interaction interferes with the creation of system-spanning shear bands and increases the yield strain. We offer
experimentally testable estimates of the lengths of nano-shear bands in the presence of the additional elements.
\end{abstract}


\section{Introduction}
\label{intro}

The history of the production of amorphous metallic glasses spans by now more than seven decades. Originally Buckel and
Hilsch \cite{54BH} accidentally formed amorphous metals by vapor deposition of one or several component systems onto an
ultra-cold substrate. Later P. Duwez \cite{60KWS} discovered that an alloy in its eutectic composition with an eutectic
temperature as low as possible (Au-Si) is very much favored to exist even at room temperature and above in a
non-crystalline form. But only the intentional addition of a third or more element made {\em bulk} metallic glasses
possible, for example A. Inoue's and W.L. Johnson's alloys in the early nineties \cite{00Ino,93PJ}.

\begin{figure}
\begin{center}
\includegraphics{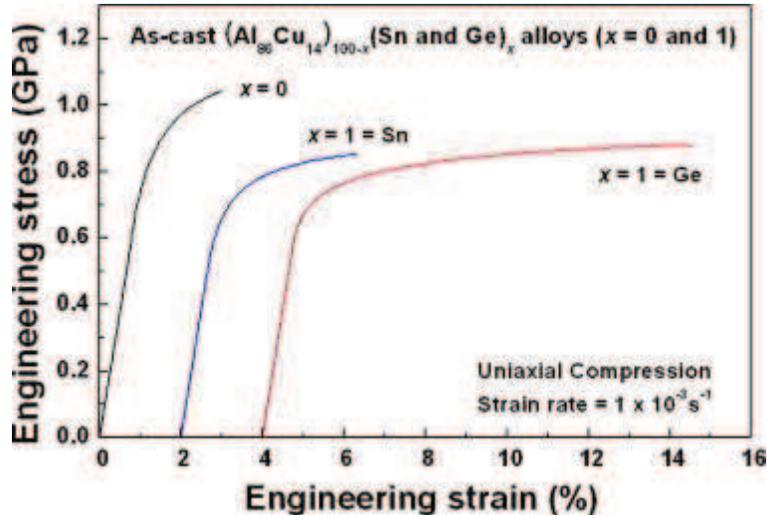}
	\caption{The effect of a small concentration of a Sn and Ge on the toughness of Al-Cu
		metallic glass.}
	\label{strength}
\end{center}
\end{figure}

For more than twenty years it has been known that small additions of a further alloy component can greatly enhance the
glass-forming-ability of metallic glasses, measured as the largest radius of a cylinder of a metallic glass without a
crystalline core. In the last ten years hundreds of microalloyed systems have been presented by the various active groups
in China, US, Japan, Korea and Europe \cite{15WLYLDL}. It was argued that an important effect of adding a minute amount of
foreign atoms is on the eutectic temperature of the mixture. There is a dip in the crystallization temperature for a liquid
system cooled down into the glass regime as a function of the microalloying concentration that reaches a minimum at some
finite but small concentration (say 1\%).
Remarkable cusps have been observed in the glass-forming-ability of many alloys with a width in composition down to 0.1\%
or even less \cite{15WLYLDL,14NDFHGJ}. These cusps reflect exactly the extremely deep lying (in temperature) eutectic
points. In other words the liquid is extremely stabilized against partitioning  into the stable crystalline or
polycrystalline phases. Whether this is due to chemical or topological barriers is an actual debate. But this
stabilization of the liquid state down to extremely low temperatures by minute additions of another component can also be
seen in the liquid properties like diffusion constants \cite{09CDUS}, formation of special local structures \cite{06SLABM}
or fragility \cite{15KSZ}. Microalloying also delays the failure of metallic glasses,
increasing its toughness as defined by the integral under the stress vs. strain curve, see Fig.~\ref{strength}.

A number of papers in the literature have proposed that the short and medium range order
in metallic glasses can be partially accounted for by icosahedra or quasi-crystalline clusters.
Of course, such quasicrystalline clusters of atoms cannot grow beyond a certain size because of
local frustration. This suggestion was originally proposed by Kivelson et al.~\cite{KK95}; this idea was extended by de
Gennes~\cite{DG02}. For example, the atomic structure of the Cu$_{35}$Zr$_{65}$ , Cu$_{50}$Zr$_{50}$ , and
Cu$_{65}$Zr$_{35}$ metallic glasses were investigated by means of high-energy X-ray diffraction and neutron diffraction,
and the geometric short-range order
found could be characterized by a variety of polyhedra~\cite{MJK09}.
Another example is the formation of an icosahedral quasicrystalline phase
followed by crystallization of tetragonal CuZr$_2$ which has been observed in the
Zr$_{70}$Cu$_{29}$Pd$_1$ glassy alloy during annealing up to 850 K~\cite{SFJ03}.

Molecular dynamics simulations have given similar clues. For example,
Lekka et al.~\cite{LIYE07} concluded from molecular dynamic
simulations of Cu$_{46}$Zr$_{54}$ glasses that 23\% of the atoms belonged to Cu-
centred icosahedral clusters and about 41\% belonged to Zr- centred
clusters. While Lee et al.~\cite{LLL08} showed that in molecular dynamics simulations of a Cu$_{65}$Zr$_{35}$ there were
polyhedral clusters, of which 15\% were ideal icosahedra.

Perhaps the most likely clusters formed in metallic glasses are the
Frank-Kasper \cite{FK58} close-packed clusters whose quasicrystalline short range
order is incompatible with either the ground state crystalline symmetry
of the surrounding matrix or the random metallic glass
phase. In some cases ~\cite{06SLABM} these form as much as
20\% of the observed clusters. Thus in Ni–P glasses up to 16\% of such
Kasper polyhedra have been found, while in Ni$_{81}$B$_{19}$ metallic glasses
Kasper polyhedra are dominant (at about 17.8\% and
7.1\%, respectively).  In fact, it has been proposed that Kasper polyhedral short
range order is the main underlying topological short range order in
metallic glasses.

In this paper we propose that the microalloying particles may be nucleating a patch of a new local structure around
themselves that frustrates the formation of the pure crystalline phase, being incompatible with both
the equilibrium crystalline order and with the glassy disorder. Whether these patches are crystalline (with different
crystal structure than the thermodynamic equilibrium phase of the unadulterated glass) or icosahedral (quasicrystalline) in
nature is still
debated but will not be important for our considerations.

Accepting the point of view that the microalloying particles organize around them a local structure
that differs from the bulk structure, we will treat them in this paper as {\em defects} in the bulk
structure, and seek a theoretical explanation for their influence on the mechanical properties of
the glass. Importantly, we will assert that on scales larger than the local patches around the foreign
elements these defects interact isotropically (i.e. the defects have spherical symmetry). We will show
that they interact with the plastic events that occur in metallic glasses under straining, and the latter
are {\em not} spherically symmetric. Rather, plastic events have quadrupolar symmetry as predicted by
the theory of Eshelby inclusions \cite{57Esh}, and see also \cite{argon}. In two dimensions the difference is easy to
characterize, in terms
of the SO(2) symmetry group; we treat the isotropic defect as having an $\ell=0$ characteristic whereas the plastic
quadrupolar displacement fields associated with plastic instabilities
have an $\ell=2$ characteristic.

In Sect.~\ref{single} we review the theory of the mechanical instability in amorphous solids that leads
to shear banding, and explain why the existence of even a {\em single} isotropic defect can defer the instability, leading
to higher toughness of the material. For the sake of clarity and simplicity we limit the discussion to systems in
2-dimensions. The generalization to 3-dimensions will be commented upon in the last section. In Sect.~\ref{density} we
treat a density of
$\ell=0$ defects and show that it is expected to result in increasing the toughness of the material, leading to the
appearance of arrested shear bands of finite length. The expected length of the arrested shear bands is
estimated in Sect.~\ref{phenom}. We also explain why an even larger concentration of foreign elements is {\em not}
beneficial for increasing the toughness of the material.  These predictions
should be tested in experiments. Finally, Sect.~\ref{summary} offers a summary, a discussion
and some comments about the road ahead.

\section{The effect of a single isotropic defect on the shear banding instability in two-dimensions}
\label{single}
In this section we construct the theoretical framework to discuss the effect of isotropic ($\ell=0$) defects
on the shear banding instability in two-dimensional amorphous solids. We begin with a short review of the
microscopic theory of the shear banding instability in the unadulterated glass.
\subsection{review of the shear banding instability in two-dimensions}

Plastic responses in amorphous solids are sensitive to temperature effects, strain rates and the mode
of external loading. For simplicity and concreteness in the present discussion we will focus on
amorphous solids in athermal conditions ($T=0$ or in practice $T\ll T_g$ where $T_g$ is the glass transition temperature).
We will also limit ourselves to pure external shear strains which are area preserving since the nature of the shear banding
instability is sensitive to loading that is non-area preserving, cf. \cite{13JGPS}. Denoting the external strain by
$\gamma$ (without tensorial indices since we have only shear strain), we consider the limit $\dot\gamma \to 0$
(quasi-static conditions) where the theory appears in its cleanest form. The numerical protocol used to produce the results shown below in Figs.~\ref{eshelby} and \ref{line} is described in the Appendix.  Under these conditions plastic instabilities are identified by the vanishing of an eigenvalue of the Hessian matrix of the system. Denoting the total energy of the system
by $U(\B r_1,\B r_2,\cdots \B r_N)$ where $\{\B r_i\}_{i=1}^N$ is the array of particle positions, the Hessian matrix
is defined by
\begin{equation}
H_{i,j} \equiv \frac{\partial^2 U(\B r_1,\B r_2,\cdots \B r_N)}{\partial \B r_i \partial \B r_j}
\end{equation}
As long as all the eigenvalues of the (symmetric and real) Hessian matrix $\B H$ are positive, the
system is stable. Plasticity is the consequence of an instability with at least one of the eigenvalues
$\lambda_p$ going to zero at some value of the strain $\gamma=\gamma_P$. It was shown that at that value
of the strain the associated eigenfunction $\Psi_p$ localizes on the particles that participate in the
plastic events \cite{12DKP}. In fact the eigenfunction which is extended as long as $\lambda_p>0$ localizes precisely
on the non-affine plastic displacement field that is associated with the plastic instability.
\begin{figure}
	\vskip 0.5 cm
	\begin{center}
		\includegraphics[scale = 0.30]{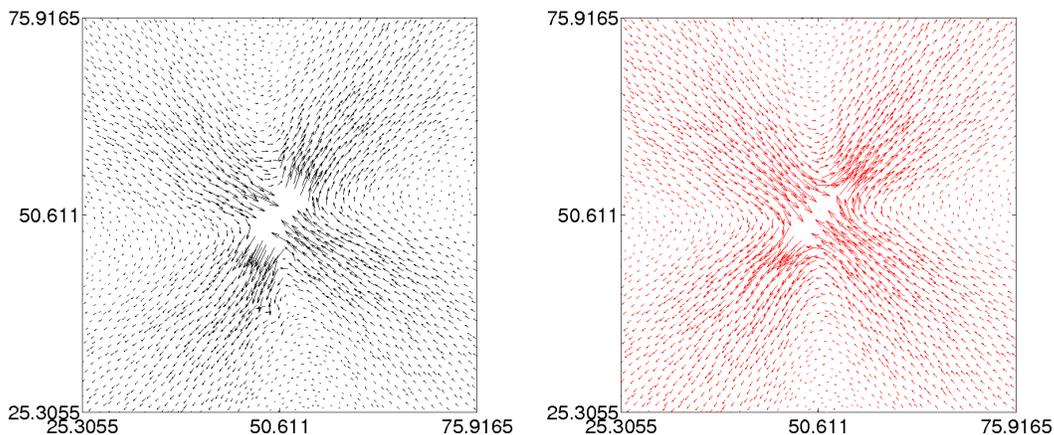}
	\end{center}
	\caption{{A window into the simulations cell with coordinates in Lennard-Jones units.} Shown is a typical plastic instability at small values of the external strain. Left panel: the displacement field associated
		with the plastic event as observed in numerical simulations, cf. Ref.~\cite{12DHP}. Right panel: the displacement field of
		an Eshelby solution where the eigenstrain and the core-size were fitted
		to the numerically found instability. The quadrupolar structure with power law decay towards infinity
		is obvious. }
	\label{eshelby}
\end{figure}
\begin{figure}
	\begin{center}
		\includegraphics[scale = 0.30]{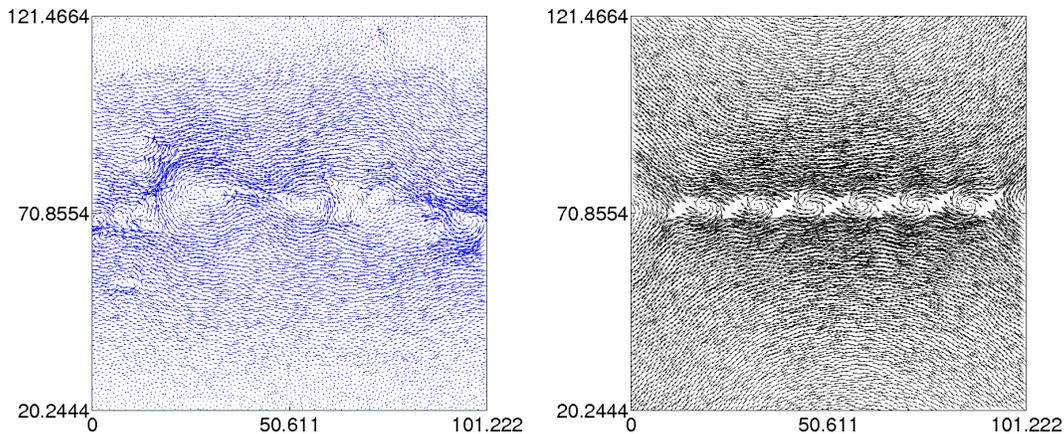}
	\end{center}
	\caption{{A window into the simulations cell with coordinates in Lennard-Jones units.} Typical plastic instability at large values of the external stain. Left pannel: the displacement field associated
		with a system-spanning plastic event, localizing the displacement around a narrow
		line \cite{12DHP}. Right panel: a model of the same event using a series of Eshelby quadrupoles. Now there
		exists a global connection between the outgoing and ingoing displacement fields on the quadrupoles, localising the
		displacement on a thin line.}
	\label{line}
\end{figure}

The nature of the displacement field associated with the plastic instability differs dramatically
when $\gamma$ is small and when $\gamma$ is large. For small values of $\gamma$, when the system
is deep in the ``elastic" region, plastic events are small (localized), having a quadrupolar structure that
is excellently modeled by the displacement field associated with an Eshelby inclusion, cf. Fig.\ref{eshelby}. At larger
values
of $\gamma$, above some theoretically computable value $\gamma_{_Y}$, the instability appears
as the simultaneous inception of a whole line of quadrupoles, system spanning, which organize the
displacement field on a narrow line with the displacement field pointing in opposite directions
above and below the line, see Fig.~\ref{line}. It was shown \cite{12DHP,13DHP} that the single quadrupole solution is a
minimal energy solution as long as $\gamma<\gamma_{_Y}$ but when $\gamma>\gamma_{_Y}$ the system-spanning line of
quadrupoles is winning, being the minimal energy solution. This solution requires
that the quadrupoles will be ``in phase", meaning that the stable (respectively unstable) direction of every
quadrupole is parallel to the stable (respectively unstable) direction of every other quadrupole.  In other words,
the quadrupoles are oriented in the same way with respect to the line joining their cores, and this line should be at
45$^0$ with respect to the principal stress axis. The reader is referred to refs.~\cite{12DHP,13JGPS,13DHP}  for a full
explanation of these findings. In the next subsection, while preparing for the analysis of the effect of the $\ell=0$
defect, we will shed more light on the phenomenon.
It is important to stress that in the quasi-static strain controlled protocols which are used in our numerical
simulations the ``strain rate" is always zero, even when system spanning plastic instabilities take place. One is
increasing the strain
infinitesimally and the system is allowed to complete its non-affine response before the strain is increased further.
\subsection{Theoretical considerations}

The elastic energy of interaction between $\ell = 0$ and $\ell = 2$ defects can be modeled as the interaction between
Eshelby inclusions, where the case of $\ell = 2$ corresponds to a purely deviatoric inclusion, and the case of $\ell = 0$
corresponds to a purely isotropic inclusion.
A calculation for the interaction between two $\ell = 2$ defects was presented in \cite{12DHP,13DHP}.
Although it is possible to extend the calculations shown in \cite{12DHP,13DHP} to include both $\ell = 0$ and $\ell =2$
defects, here we adopt a different approach whose elaboration can be found in Refs.~\cite{Moshe2015PNAS,Moshe2015PRE}.
This approach relies on a geometric formulation of incompatible elasticity. According to this approach localized plastic
deformations are sources for residual stresses, whose energetic implications are very similar to those of charge densities
in electrostatics.
These stress densities are quantified by a single scalar function $\bar{K}$, and in the case of an elastic solid they act
as source terms for the Airy stress function $\psi$ which solves the bi-Laplace equation
\begin{equation}
\Delta\Delta \psi(\mathbf{x}) = Y \bar{K}(\mathbf{x}),
\label{eq:Biharmonic}
\end{equation}
where $\Delta$ is the Laplace operator and $Y$ is the Young's modulus.

For example, the far field description of an $\ell=0$ defect within this formulation is given by an elastic charge
singularity of the form
\begin{equation}
\bar{K}(\mathbf{x}) = \frac{1}{2} P \Delta \delta(\mathbf{x}).
\end{equation}
Here $P=\pi a_{\rm iso}^2 \epsilon_{\rm iso}^*$ where $a_{\rm iso}$ will take the physical meaning the core size of the
isotropic ``patch"
around our microalloying particle and $\epsilon_{\rm iso}^*$ is the eigengstrain associated with  the isotropic inclusion.
Positive or negative values of $P$ correspond respectively to expansion or contraction due to the inclusion.

In electrostatics, two singular charge densities $\rho_1$ and $\rho_2$ induce electrostatic potentials $\phi_1$ and
$\phi_2$ through the Poisson's equation. The interaction between two such electric charge densities is $U_{el} = \int
\phi_1(\mathbf{x}) \rho_2(\mathbf{x})\,  d\mathbf{x} $.
In \cite{Moshe2015PRE}  it was shown that the same holds for elastic stress densities. That is, consider two singular
stress densities described by $\bar{K}_1$ and $\bar{K}_2$. Each of these induces (through Eq. \ref{eq:Biharmonic}) an
elastic potential denoted as $\psi_1$ and $\psi_2$. The energy associated with their interaction is
\begin{equation}
U = \int \psi_1(\mathbf{x}) \bar{K}_2(\mathbf{x}) \, d\mathbf{x}=\int \psi_2(\mathbf{x}) \bar{K}_1(\mathbf{x}) \,
d\mathbf{x} \ .
\label{eq:Interaction}
\end{equation}
This result is the key for obtaining simple explicit expressions for the interactions between defects, and their
interactions with external fields.

A short list of possible singularities $\bar K$ together with their corresponding elastic potentials $\psi$ are listed in
Table \ref{tab:table1}.

\begin{table}[h]
	\begin{center}
		\begin{tabular}{lcc}
			Type   &  $\bar{K}$ & $\psi$  \\ \hline
			Point &
			$ \frac{1}{2} P \,\Delta \delta(\xvec)$
			&
			$(Y P \,/4\pi) \ln|\xvec|$
			\\
			Quadrupole &
			$\frac14 (\nabla^T\cdot \Qvec\cdot \nabla) \delta(\xvec)$
			&
			$(Y/16\pi)\, ({\hat{\xvec}^T}\cdot\Qvec\cdot\hat{\xvec} )$
			\\
			External stress&
			--
			&
			$\frac12 (\xvec^T \cdot \Cof{\sigma} \cdot \xvec)$
			\\
		\end{tabular}
		\caption{Elastic charge sources $\bar{K}$ and their corresponding elastic potentials $\psi$. The superscript
			``T" refers to ``transpose".}
		\label{tab:table1}
	\end{center}
\end{table}

Here $\nabla$ is the nabla operator,
$\Qvec$ is a traceless symmetric matrix
\begin{equation}
\Qvec = \matrixII{Q_1}{Q_2}{Q_2}{- Q_1},
\end{equation}
and $\Cof{\cdot}$ is the Cofactor of the matrix concerned.

Given a purely deviatoric inclusion oriented with angle $\theta$ measured from the $x$ axis, its quadrupolar charge is
\begin{equation}
\Qvec = Q \matrixII{\cos 2\theta}{\sin 2\theta}{\sin 2\theta}{- \cos 2\theta}.
\end{equation}
Here $Q=\pi a_{\rm quad}^2 \epsilon^*_{\rm quad}$ with, as before $a_{\rm quad}$ being the core size of the
quadrupolar singularity and $\epsilon^*_{\rm quad}$ its eigenstrain.
Given {\em two} such deviatoric inclusions with magnitudes $Q_1$ and $Q_2$, oriented by $\theta_1$ and $\theta_2$ measured
from the line connecting them, their elastic interaction energy is obtained by substituting $\psi$ of one quadrupole and
$\bar{K}$ of the second quadrupole in Eq.~(\ref{eq:Interaction}):
\begin{equation}
U_{QQ} = \frac{Y Q_1 Q_2}{16 \pi r^2} \cos(2 \theta_1 + 2\theta_2).
\label{qq}
\end{equation}

Similarly, we obtain an expression for the interaction between an $\ell = 0$ defect located at the origin, with an $\ell =
2$ defect located at distance $r$, with an orientation $\theta$ measured from their connecting line
\begin{equation}
U_{QP} = \frac{Y Q  P}{8 \pi r^2} \cos(2 \theta).
\label{qp}
\end{equation}

With an external fields present, its interaction with the quadrupoles should be taken into account. Using the same
procedure (see for example \cite{Moshe2015PNAS})  we find that the interaction of an external shear $\sigma$ with a
quadrupole is
\begin{equation}
U_{Q\gamma} = -\frac{1}{2} Q Y \gamma \sin(2 \theta)
\label{qsig}
\end{equation}
where now $\theta$ is measured from the direction of the shear's principal axis, taken below to be the $x$ axis.
Here $\gamma$ is the strain measured with respect to an appropriate reference point. Note that external fields do {\em not}
interact with $\ell=0$ defects.

\subsection{Formation of shear band without microalloying}

In previous work it was shown that when the external strain is sufficiently large, a line of quadrupoles cab appear
spontaneously as a results of plastic instability. Their orientations appear highly correlated.
To see this using the present formalism deduce from Eq.~(\ref{qsig}) that the energy is minimized for $\theta=\pi/4$ (for
positive strain).

\begin{figure}
	\begin{center}
		\includegraphics[width=0.45\columnwidth]{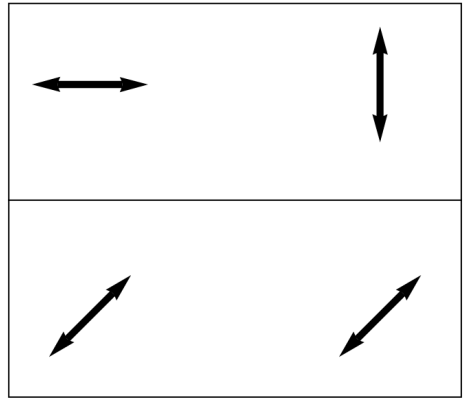}
	\end{center}
	\caption{Possible relative orientations of two quadrupoles that minimize their interaction energy without
		an external strain. With external shear strain the degeneracy is removed, and only the lower panel survives.}
	\label{quad}
\end{figure}

To explain the formation of the quadrupoles along a line we examine their interaction with each other,
rewriting Eq.~(\ref{qq}) for a general orientation of the line with respect to the $x$ axis:
\begin{equation}
U_\text{QQ} = \frac{Y Q^2}{16 \pi r^2}\cos(2\theta_1+2\theta_2 - 4\phi)\ ,
\label{qq1}
\end{equation}
where $\phi$ is the angle between the $x$ axis and the line that connects the quadrupoles.
This interaction energy is minimized for $2\theta_1+2\theta_2 - 4\phi=\pi$. In the absence of an external field there is a
degeneracy in the energy minimizing configurations. Two representative energy
minimizing configurations are shown in Fig.\ref{quad}.
When an external shear exists, the energy is minimized for $\phi=0$, that is the quadrupoles are aligned parallel to the
external shear, all with the same angle of $\pi/4$.
These results hold for any set of quadrupoles; given $N$ quadrupoles subjected to external shear, their optimal position is
along a line with the same orientation of $\pi/4$. Arranged like this they form a connection
between the incoming and outgoing directions of the quadrupolar displacement field, resulting in a shear band.
It was shown in Refs.~\cite{12DHP,13DHP} that at sufficiently large external strain $\gamma\ge\gamma_{_{\rm Y}}$
this solution is energetically favorable compared to a single localized quadrupolar displacement field. We will explain now
that in the presence of isotropic defects one must increase $\gamma$ further to allow for the
spontaneous appearance of this system-spanning instability.

\subsection{The effect of an isotropic defect}

An $\ell=0$ defect describes the uniform expansion or contraction of a small region inside a material. We rewrite now
Eq.~(\ref{qp}) for a general orientation
\begin{equation}
U_\text{QP} = \frac{Y P Q}{8 \pi r^2} \cos(2\theta - 2\phi) \ ,
\end{equation}
where the defect is at the origin and the quadrupole is at a point $(r,\phi)$. The angle $\theta$ is the quadrupole
orientation.

\begin{figure}
	\begin{center}
		\includegraphics[width=0.45\columnwidth]{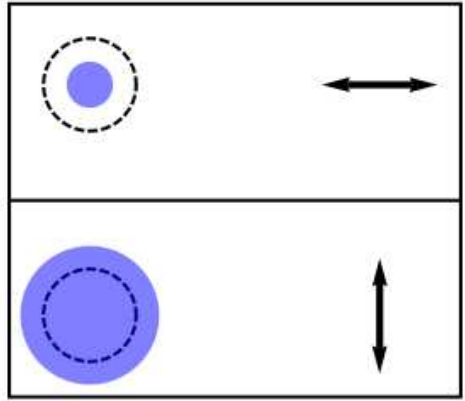}
	\end{center}
	\caption{The orientations of a quadrupole that minimize its interaction energy with
		an $\ell=0$ defect. The two orientations are associated with a negative or a positive $P$.}
	\label{PQ}
\end{figure}

Now the value of $\theta$ that minimizes the energy depends on the sign of $P$. In Fig.~\ref{PQ} we show two different
optimal relative configurations of the quadrupole associated with opposite signs of $P$.
One should note that it is {\bf not} the size of the microalloying atom which is at stake here, but
rather the contraction or expansion effect of the organized patch around the atom. Much work had been
devoted to analyzing ``small" or ``large" atoms \cite{ZQPCGZE12}, but in reality the important physics lies
in their ability to nucleate patches rather than their own size.
In these notes we assume a contraction ($P<0$), hence the minimizer is obtained for
\begin{equation}
\theta + \phi=\pi/2 \ ,
\end{equation}
meaning that the quadrupole is perpendicular to the connecting line of the quadrupole and the isotropic defect.

\subsection{Array of quadrupoles near a single isotropic defect}

In this subsection, which is central to our theory, we demonstrate how the existence of a single
isotropic defect interfere with the creation of a shear band via a system spanning plastic instability

Consider a linear array of $2N$ quadrupoles with charges $Q$, positioned at $(x_i,y_i) = ((2i + 1)L,r)$, having
orientations $\theta_i$, with $i\in(-N,-N+1,...,N-2,N-1)$.
At the origin there is an isotropic defect of strength $P$, and the system is subjected to external strain.
The elastic interaction energy is
\begin{eqnarray}
\label{inten0}
&&U= -\frac{1}{2} Y \gamma Q \sum_{i} \sin(2\theta_i) + \sum_{i\neq j}\frac{E Q^2}{16 \pi r_{ij}^2} \cos(2\theta_i +
2\theta_j)\nonumber\\&& + \frac{Y P Q}{8\pi} \sum_i \brk{\frac{x_i^2-y_i^2}{(x_i^2+y_i^2)^2} \cos(2\theta_i)+\frac{2 x_i
		y_i}{(x_i^2+y_i^2)^2} \sin(2\theta_i)} \nonumber\\&& + U_S\ .
\end{eqnarray}
where $\B r_{ij}$ are the vector distances between the quadrupoles $i$ and $j$ and $U_S$ is composed of the self energies
of the quadrupoles and the isotropic defect.

At this point we seek the orientation of the set of quadrupoles that minimizes the total elastic energy. For $P=0$ the
problem is reduced back to the one studied in \cite{12DHP,13DHP}. The result in this case is simply $\theta_i=\pi/4$.

\begin{figure}[h]
	\begin{center}
		\includegraphics[scale = 0.35]{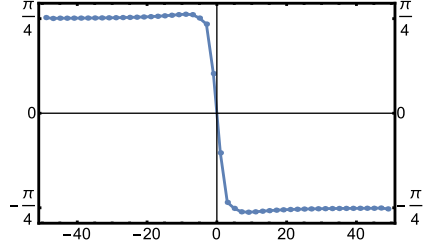}
	\end{center}
	\caption{The angle of the quadrupole $\theta$ of the set of 50 quadrupoles in the presence of an isotropic defect $\hat p =
		5$, $\zeta = 5$, and no shear $\Gamma=0$.}
	\label{optimal}
\end{figure}

In the presence of an isotropic defect the problem cannot be solved analytically, and we resort to a numerical solution.
There are three independent dimensionless parameters in Eq.~(\ref{inten0}). The first is the ratio
of amplitudes of the isotropic and quadrupolar defect $\hat p = P/Q$. The second is the scaled distance $\zeta= r/L$ and
last is the rescaled external strain $\Gamma = \gamma \frac{L^2}{Q}$. In the following figures we plot the optimal
orientations of quadrupoles for several values of $\hat p,\zeta,\Gamma$.
In Fig.~\ref{optimal} we show the optimal orientation of a set of $50$ quadrupoles in the presence of an isotropic defect
with $\hat p = 5$, $\zeta = 5$, and no external strain, i.e. $\Gamma=0$.

The $x$ axis in this figure corresponds to the position of the quadrupole along the line, and the $y$ axis corresponds to
the angle $\theta$ of the direction of the quadrupole that minimized the elastic energy. Note that when there is no
external strain, there are two equal orientation $\theta=\pm \pi/4$ for which the energy of the line of quadrupoles is at a
minimum. When an istorpic defect is placed at the origin, this can cause a flip from
$\theta= \pi/4$ to $\theta=- \pi/4$. We should stress however that at $\gamma=0$ one does not expect
a line of quadrupoles, since the single quadrupole solution has a lower energy.

\begin{figure}[h]
	\begin{center}
		\includegraphics[scale = 0.23]{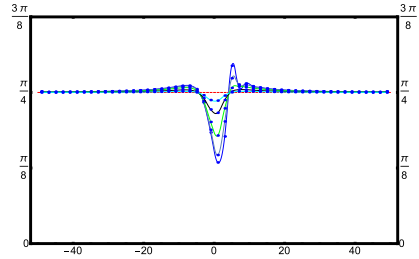}
	\end{center}
	\caption{The angle of the quadrupole $\theta$ of the set of 50 quadrupoles in the presence of an isotropic defect $\hat p =
		5$, $\zeta = 5$, and for various shear values: $\Gamma=5 \times 10^{-3}, 10^{-2},3\times 10^{-2},10^{-1},3\times 10^{-1})$.
		As the external strain is increased the effect of the perturbation decreases.}
	\label{line2}
\end{figure}

In Fig.~\ref{line2} we plot a similar graph, for several non-zero external shear strains $\Gamma =(5 \times 10^{-3},
10^{-2},3\times 10^{-2},10^{-1},3\times 10^{-1})$.
With a finite external strain the preferred angle $\theta$ sufficiently far from the $\ell=0$ defect is
always $\theta=\pi/4$. The presence of the defect perturbs this angle, with the quadrupoles that are
closest to the defect being disturbed most. Note the asymmetry in the amount of perturbations of the
angles of the quadrupoles on the right and on the left parts of the isotropic defect. This stems from the fact that the
{\em positions} of the quadrupoles are right-left symmetric, their orientation is not.
The interactions with quadrupoles on the right and and on the left are not the same.

Of course, with the quadrupoles perturbed, the associated
displacement field cannot be described by a thin system spanning shear-band. In order to obtain a spanning shear-banding
instability one needs to increase the external strain to overcome the perturbation of the
isotropic defect. While this does not provide yet a full explanation to the
data shown in Fig.~\ref{strength}, it indicates the physical origin of the phenomenon. To understand
it fully we need to consider a density of isotropic defects as done in the next section.

\begin{figure}[h]
	\begin{center}
		\includegraphics[scale = 0.23]{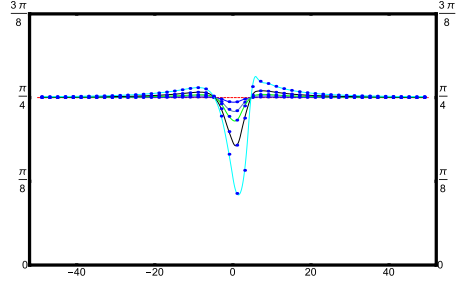}
	\end{center}
	\caption{ The angle of the quadrupole $\theta$ of the set of 50 quadruoples in the presence of an isotropic defect
		$\Gamma=0.1$, $\zeta = 5$, and for various values of $\hat p$: $\hat p=1,3,5,10,20$. As $\hat p$ is increased
		the effect of the $\ell=0$ defect strengthens.}
	\label{line3}
\end{figure}

\begin{figure}[h]
	\begin{center}
		\includegraphics[scale = 0.23]{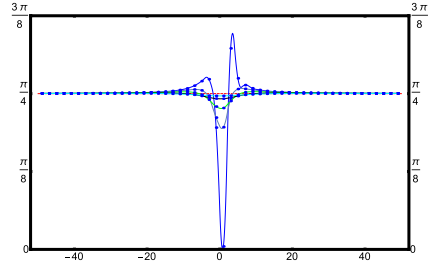}
	\end{center}
	\caption{The angle of the quadrupole $\theta$ of the set of 50 quadrupoles in the presence of an isotropic defect $\hat
		p=5$, $\Gamma=0.1$, and for various values of $\zeta$: $\zeta=2,4,6,10,15$. As $\zeta$ increases
		the effect of the $\ell=0$ defect decreases.}
	\label{line4}
\end{figure}

For completeness we show results of a similar calculation for a given value of the external strain but with other values of
$\hat p$ and $\zeta$. In Fig.~\ref{line3} we take $\zeta=5$, $\Gamma=0.1$, and $\hat p = (1,3,5,10,20)$.
In Fig.~\ref{line4} we take $\hat p=5$, $\Gamma=0.1$, and $\zeta=(2,4,6,10,15)$

The reader should note the interesting up-down-up oscillation in the response of the computed angles
to the presence of the isotropic defect. The upward oscillation can be interpreted as a screening effect.

\section{The increase in yield strain due to a density of isotropic defects}
\label{density}
In this section we attempt to demonstrate the increase in the yield strain due to the existence
of a density of $\ell=0$ defects. To this aim we consider below an infinite two dimensional body containing a dilute array
of isotropic defects. As a preparatory step
we consider first a square array of isotropic defects of fixed strength $P$. Understanding this easier
case will help later with the case of a random distribution of isotropic defects.

Denote as $l_p$ the distance between neighboring isotropic defects. With $(i,j)\in \mathbb{N} \times \mathbb{N} $ their
positions in cartesian coordinates are \begin{equation}
(x_i,y_j) = ((\alpha  + i) \cdot l_p,  (\beta  + j) \cdot l_p )\ , \quad 0<\alpha,\beta<1 \ .
\end{equation}
The material is subjected to external shear strain $\gamma$, and the elastic energy stored in the system at this stage is
set to zero. At this point we ask if the formation of a linear array of quadrupoles can reduce the elastic energy stored in
the system.

We start with a single quadrupole located at the origin. This quadrupole interact with the external shear and with the
isotropic defects. In addition it contributes a self interacting term to the elastic energy. The interaction energy of such
quadrupole with an isotropic defect located at $(x_0,y_0)$ is
\begin{equation}
E_{qp} = \frac{ P Q Y }{2 \pi} \brk{ \frac{x_0^2 - y_0^2}{\brk{x_0^2 + y_0^2}^2} \cos 2\theta + \frac{2 x_0 y_0}{\brk{x_0^2
			+ y_0^2}^2} \sin 2\theta}
\end{equation}
Substituting the positions of the isotropic defects and summing over all of them we find the elastic energy interaction
\begin{equation}
E^{i}_{qp} = \frac{ P Q Y }{2 \pi} \brk{ f(\alpha,\beta) \cos 2\theta_i + g(\alpha,\beta) \sin 2\theta_i},
\end{equation}
where
\begin{equation}
f(\alpha,\beta) = \frac{1}{l_p^2}\sum_{i=-\infty}^{\infty} \sum_{j=-\infty}^{\infty}\frac{x_i^2 - y_j^2}{\brk{x_i^2 +
		y_j^2}^2} \equiv \frac{1}{l_p^2} \tilde{f}(\alpha,\beta),
\end{equation}
and
\begin{equation}
g(\alpha,\beta) = \frac{1}{l_p^2}\sum_{i=-\infty}^{\infty} \sum_{j=-\infty}^{\infty}\frac{2 x_i y_j}{\brk{x_i^2 +
		y_j^2}^2}\equiv \frac{1}{l_p^2} \tilde{g}(\alpha,\beta).
\end{equation}
Note that for the case of a random distribution of defects $f$ and $g$ are different for each quadrupole, and therefore
effectively $\alpha$ and $\beta$ are functions of quadrupoles positions. We will use this later.

The interaction with the external shear is
\begin{equation}
E_\text{\rm Shear} = - \frac{1}{2} Y Q \gamma \sin 2\theta \ ,
\end{equation}
and the self interaction term is
\begin{equation}
E_\text{\rm Self} = \frac{Y Q^2}{16 \pi a^2} \ .
\end{equation}

Since we are interested in a linear array of quadrupoles we consider a domain in the material $(x,y) \in
[0,L]\times[-\infty,\infty]$, with a set of quadrupoles located at $(x_i,y_i) = (i \cdot l_q, 0)$. Here $l_q$ is the
distance between the quadrupoles. We further simplify the problem by assuming $L = k \cdot l_q =n \cdot l_p $, that is the
length $L$ is equal to an integer multiple of $l_p$, and so is $l_q = m \cdot l_p$.

We denote by $\rho_p \equiv \frac{n}{L}=\frac{1}{l_p}$ the isotropic defects density, and  $\rho_q \equiv
\frac{k}{L}=\frac{1}{l_q}$ the quadrupoles density.
In addition to the list of interactions mentioned above, each quadrupole interacts now with all the other quadrupoles. This
interaction energy of the $i$'th quadrupole with all the others is
\begin{equation}
E^i_{qq} = \sum_{j\neq i}^{}\frac{Y Q^2}{16 \pi R_{ij}^2} \cos \brk{2 \theta_i + 2\theta_j}
\end{equation}
Substituting the quadrupoles locations we find
\begin{equation}
E^i_{qq} = \frac{Y Q^2}{16 \pi l_q^2} \sum_{j=-\infty}^{\infty} \frac{\cos \brk{2 \theta_i + 2\theta_j}}{(i-j)^2}
\end{equation}

We will use now all the results above for the calculation of the slope of the energy per unit length, vs. quadrupoles
density.

We have $ k$ quadrupoles in the considered domain. Therefore the elastic energy stored in the domain is
\begin{equation}
E_L =   \sum_{i=1}^{k} E_{\rm Self} + E^i_{pq} + E_\text{\rm Shear}^i + E^i_{qq}
\end{equation}
The total energy stored in the system is $E_\text{tot} = \frac{L_\infty}{L} \cdot E_L$, where $L_\infty$ is the size of the
system which goes to infinity. The energy per unit length is therefore
\begin{eqnarray}
&&\frac{E_\text{tot}}{L_\infty} = \frac{k}{L} \cdot \frac{Y Q^2}{16 \pi a^2} \\ &&+ \frac{ P Q Y }{2 \pi L}
\sum_{i=1}^{k}\brk{ f(\alpha,\beta) \cos 2\theta_i + g(\alpha,\beta) \sin 2\theta_i} - \nonumber\\&&\frac{Y Q \gamma}{L}
\frac{1}{L}\sum_{i=1}^{k}\sin 2\theta_i + \sum_{i=1}^{k} E^i_{qq} \ . \nonumber
\end{eqnarray}
Using the following abbreviation
\begin{equation}
\left<\dots\right> = \frac{1}{k}\sum_{i=1}^{k}
\end{equation}
the above equation reads
\begin{eqnarray}
&&\frac{E_\text{tot}}{L_\infty} = \frac{k}{L} \cdot \frac{Y Q^2}{16 \pi a^2} \\&& + \frac{ P Q Y }{2 \pi L} k \brk{
	f(\alpha,\beta) \left<\cos 2\theta_i\right> + g(\alpha,\beta) \left<\sin 2\theta_i\right>}\nonumber\\&& - \frac{Y Q \gamma
	k}{L} \left<\sin 2\theta_i\right> + \frac{1}{L} \sum_{i=1}^{k} E^i_{qq} \ . \nonumber
\end{eqnarray}
The last term in this expression is
\begin{eqnarray}
&&\frac{1}{L} \sum_{i=1}^{k} E^i_{qq} = \frac{k}{L} \cdot \frac{Y Q^2}{16 \pi l_p^2} \left< \sum_{j=-\infty}^{\infty}
\frac{\cos \brk{2 \theta_i + 2\theta_j}}{(i-j)^2}\right> \nonumber\\&&=  \frac{Y Q^2}{16 \pi} \rho_q^3 \left<
\sum_{j=-\infty}^{\infty} \frac{\cos \brk{2 \theta_i + 2\theta_j}}{(i-j)^2}\right>
\end{eqnarray}
Since this term scales like $\rho_q^3$ it does not contribute to the slope at $\rho_q = 0$. Therefore the first three terms
in $\frac{E_\text{tot}}{L_\infty}$, which are linear in $\rho_q$, are the only relevant terms for this problem.
\begin{eqnarray}
&&\left(\frac{\partial}{\partial \rho_q} \frac{E_\text{tot}}{L_\infty}\right)_{\rho_q = 0} = \frac{Y Q^2}{16 \pi a^2}
\nonumber\\&&+ \frac{ P Q Y }{2 \pi} \brk{ f(\alpha,\beta) \left<\cos 2\theta_i\right> + g(\alpha,\beta) \left<\sin
	2\theta_i\right>} \nonumber\\&&- {Y Q \gamma} \left<\sin 2\theta_i\right> \ .
\end{eqnarray}
Defining
\begin{equation}
\gamma_c \equiv \frac{Q}{16 \pi a^2} \frac{1}{\left< \sin 2 \theta \right>} + f(\alpha,\beta) \frac{ P }{2 \pi}
\frac{\left< \cos 2 \theta \right>}{\left< \sin 2 \theta \right>} + g(\alpha,\beta) \frac{ P }{2 \pi}
\end{equation}
we get
\begin{eqnarray}
&&\Big(\frac{\partial}{\partial \rho_q} \frac{E_\text{tot}}{L_\infty}\Big)_{\rho_q = 0} = \Big(\frac{Y Q^2}{16 \pi a^2} +
\frac{ P Q Y }{2 \pi}  \big(f(\alpha,\beta) \left<\cos 2\theta_i\right> + \nonumber\\&&g(\alpha,\beta) \left<\sin
2\theta_i\right>\big)\Big) \times \Big(1 - \frac{ \gamma}{\gamma_c}\Big) \ .
\end{eqnarray}

This is the main result of this section. We will now explain its implications for the problem at hand.

The defects magnitude $P$ has the dimensions of squared length. We therefore define the dimensionless parameter $\eta =
\frac{P}{l_p^2}$ as the dilution of the isotropic defects.

Start with the limit $\eta\to 0$ where there are no isotropic defects. This limit is achieved equivalently using $P \to 0$
or $l_p \to \infty$. In this limit
\begin{equation}
\gamma_c \equiv \frac{Q}{16 \pi a^2} \frac{1}{\left< \sin 2 \theta \right>}.
\end{equation}
In the absence of isotropic defects the energy is minimized for $\theta_i = \pi/4$, hence
\begin{equation}
\gamma_c \equiv \frac{Q}{16 \pi a^2},
\end{equation}
which is consistent with the result of Ref.~\cite{12DHP}.

Next we consider the limit of small nonzero dilution $|\eta| \ll 1$.
In this limit the energy minimizing orientations $\theta_i$ deviate slightly from $\pi/4$.
We find
\begin{equation}
\sin 2\theta_i \approx 1 - \eta^2 \delta_i^2 \ ,
\end{equation}
\begin{equation}
\cos 2\theta_i \approx  - \eta \delta_i \ .
\end{equation}
Expanding $\gamma_c$ in powers of $\eta$ we find
\begin{equation}\gamma_c = \gamma_c^0 + \frac{\tilde{g}(\alpha,\beta) - 2 \tilde{f}(\alpha,\beta) \left< \delta_i \right>
}{2\pi} \eta +O(\eta^2)\end{equation}
This shows that the critical value of the external shear may increase or decrease. We now explain the relevance of these
finding to the increase in the toughness of the material.

\begin{figure}
\begin{center}
	\includegraphics[width=0.75\columnwidth]{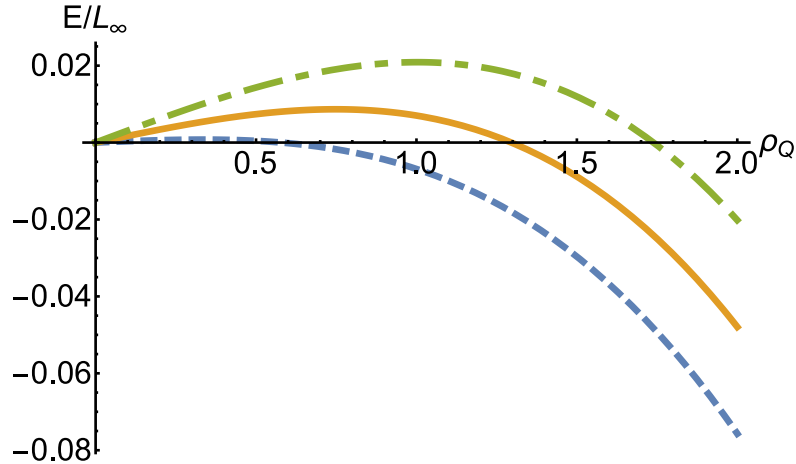}
	\caption{Energy density as a function of quadrupole density. The yellow continuous line corresponds to $(\alpha,\beta)
		= (0.85,0.15)$. The blue dashed line is for $(\alpha,\beta) = (0.15,0.5)$. The green dashed
		dotted line is for $(\alpha,\beta) = (0.15,0.15)$. The parameters used for this plot are: $P/lp^2 = -0.01$,$Q = 0.158$. The
		external shear strain is the yield strain corresponding to $P=0$, $\gamma = Q/(16 \pi a^2)$. }
\end{center}
\label{EvsRAlphaBeta}
\end{figure}

In the presented setup $\alpha$ and $\beta$ are determined by energy minimization, which in turn determine
$\tilde{f},\tilde{g}$. The functions $f$ and $g$ diverge for values of $\alpha$ and $\beta$ that correspond to the
locations of the isotropic defects. Since these defects have a size, the divergence of $f$ and $g$ has a natural  cutoff.
In fact the isotropic defects are randomly distributed rather then forming an ordered array. An effective way of taking
this into account is to take $\alpha$ and $\beta$ to be random functions of the quadrupole positions - describing the
random location of each quadrupole with respect to the isotropic defects structure. In this case the slope that we
calculated is also position dependent - that is adding a single quadrupole at some position may be very expensive
energetically, while for another position it can reduce the energy.
Accordingly,  while some values of $\alpha$ and $\beta$ encourage the formation of quadrupoles, other values suppress it.
In Fig.~\ref{EvsRAlphaBeta} we plot the energy density as a function of quadrupole density for different values of $\alpha$
and $\beta$.

In order to prevent the formation of a macroscopic shear band it is enough to suppress the formation of quadrupoles at some
region. This will prevent the formation of a system-spanning array of quadrupoles.
The fact that the slope for the energy associated with  the formation of quadrupoles become positive at some regions shows
that a random distribution of isotropic defects prevent the formation of system-spanning array of quadrupoles.

We therefore predict that a random distribution of isotropic defects allow for the formation of {\em finite} linear arrays
of quadrupoles, whose lengths are determined by the density and magnitudes of the isotropic defects, and by the magnitude
of the external shear. For large enough, but finite, external shear the mean length of a linear array of quadrupoles will
reach the system size and a system spanning shear-band will form.

\begin{figure}
\begin{center}
\includegraphics[width=0.75\columnwidth]{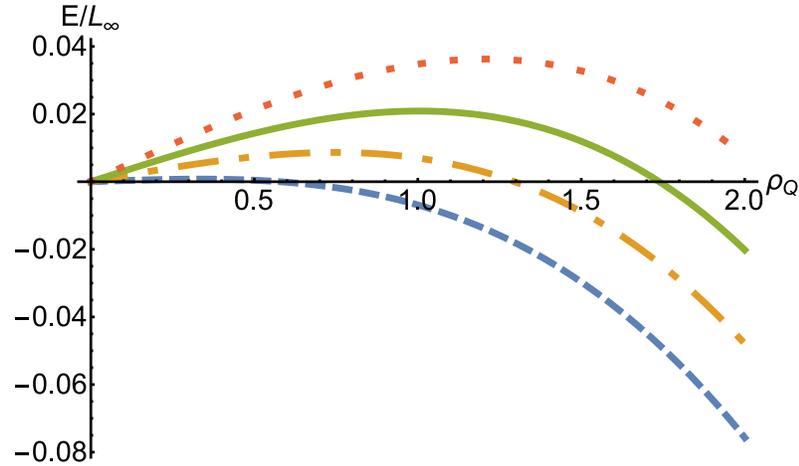}
	\caption{Energy density as a function of quadrupoles density for several values of $\eta$, and for $(\alpha,\beta) =
		(0.15,0.85)$. All parameters are fixed. The external shear corresponds to the yield strain for $\eta=0$. Increasing
		$|\eta|$ result with a larger slope, hence there are regions for which the formation of quadrupoles is energetically
		unfavorable.  }
        \end{center}
	\label{EvsRPs}
\end{figure}

It is now left to show that by increasing $\eta$ the slope of the energy density vs. quadrupole density, increases. In
Fig.~\ref{EvsRPs} we plot the energy density as a function of the quadrupole density for several values of $\eta$.

\begin{figure}
\begin{center}
\includegraphics[width=0.75\columnwidth]{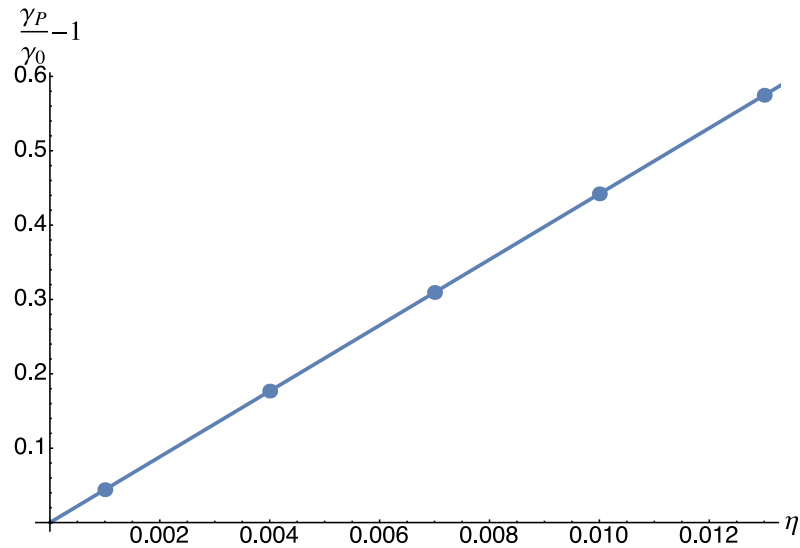}
	\caption{The percentage change in yield strain at $\rho_q = 0$, for several values of $\eta = P/l_p^2$. For $\eta=0$
		there is no change in yield strain}
\end{center}
	\label{GammaVsP}
\end{figure}

Finally we want to find quantitatively how the toughness of the material is improved by increasing $\eta$. In
Fig.~\ref{GammaVsP} we plot the percentage change of yield strain, as a function of $\eta$.
We see that for small values of $\eta$ the effect is linear in $\eta$. For these small values of $\eta$ we can get an
increase of $60$\% in yield strain.
These results are depend also on the core size of the isotropic defects. We chose $(\alpha,\beta)$ to be as close as
possible to one isotropic defect. For smaller cutoffs the increase in yield strain will be larger. In the discussion
section we will explain that for much larger values of $\eta$ the effect reverses and finally disappears.

In summary, we showed that the presence of isotropic defects may encourage or discourage locally the formation of
quadrupoles. Since a shear-band requires the formation of a system spanning array of quadrupoles, it is enough to focus on
the neighborhoods in which quadrupoles formation is suppressed.
The formation of a system spanning linear array of quadrupoles will always be frustrated in regions that are close to an
isotropic defect that happen to increase the interaction energy.
A random distribution of isotropic defects will result in regions for which the slope of the energy density becomes
positive and large. For a dilute distribution of isotropic defects the slope of the energy density may increase up to 60
percents, preventing the formation of system-spanning arrays of quadrupoles.
The mechanism described here is independent of the sign of the charge of the isotropic defects. The
upshot of this discussion is that we expect to find in such materials ``nano-shearband" whose lengths are determined by the
isotropic defects distribution, and by the external shear $\gamma$.

\section{Estimating the length of arrested shear bands}
\label{phenom}

In this section we examine the consequences of having a percentage $c$ of microalloying particles in
the amorphous material. Let us assume that each such particle organizes a patch around it involving
$n\sim O(10^2)$ other particles of the surrounding material. Then a fraction $f$ of the material where
\begin{equation}
f=nc \ ,
\end{equation}
will consist of such patches. As at higher concentrations patches can overlap, the associated volume
fraction of organized material $\phi$ will be
\begin{equation}
\phi = 1-\exp{-f} \ .
\end{equation}
It is likely that the effect of the microalloying elements would be at its maximum when the patches
begin to percolate and span the system. By ``percolation" here we assume the universality class
of 2-dimensional continuum percolation with circular inclusions. In this class the critical value
for percolation is $\phi_c\approx 0.67$ and $f_c\approx 1.13$ \cite{90LT}. This agrees with the order of
magnitude of $c_c\approx 1\%$ which is a commonly quoted efficacious percentage for  microalloying particles.
For higher concentrations the effect of the microalloying particles can no longer be considered as
``defects". Their surrounding start to become the whole material, and our discussion above stops being
relevant.

Let us first consider $c\ll c_c$. In this regime the typical distance between microalloying particles
and plastic quadrupoles is $\xi$,
\begin{equation}
\xi = c^{-1/2} \lambda \ ,
\end{equation}
where $\lambda$ is the typical interatomic distance. In comparison, the radius of a single patch is
about $a_{\rm iso}\approx \sqrt{n} \lambda \approx 10 \lambda$. Thus at very low values of $c$, $\xi \gg a_{\rm iso}$.

At this point consider such a density of $\ell=0$ defects and a strained system that is attempting
to create a shear band by aligning quadrupoles along a single line. With random distribution of the
defects we expect a typical distance to the line of defects to be of the order of $\xi$. Taking
into account the perturbation effect of the $\ell=0$ defect on the shear banding instability
discussed above, we can expect that no nano shear-band will form longer than a length of the order of $\xi$.
With the numbers considered above we expect the nano shear-bands to have a length of the order of
150-500 Angstroms. We note that this prediction can be put to direct experimental test by changing $c$ and measuring the
observed length of micro shear bands in the material.

With increasing the percentage $c$ of microalloying particles, one expects that all the material
parameters will change, including density, conductivities, the elastic moduli and the yield stress. Though
an exact calculation of the dependence of all these material parameters on $c$ is a complex problem, one can make a simple
mean field estimate by applying the rule of mixtures to the particular material property of interest. For example, let us
consider the shear modulus $\mu$ and its dependence on $c$ in a block of material under uniaxial tension $\sigma$. We will
need to specify the shear modulus of the the background material in the absence of microalloying to be $\mu_m$ and the
shear modulus of the $\ell=0$ inclusions to be $\mu_i$ (naturally, depending on the material involved, $\mu_i >\mu_m$ or
$\mu_i <\mu_m$). Also as the stress is homogenous we expect $\sigma_m = \sigma_i = \sigma$. As the total strain in the
material is $\epsilon = \phi \epsilon_i + (1-\phi ) \epsilon_m$ we find
\begin{equation}
\mu (c) = \frac{1}{(1-\exp{-n c})/\mu_i + \exp{-n c}/\mu_m}.
\end{equation}
Similar arguments can be made for for thermal and electrical conductivities. A particular simple case is the mass density
$\rho = \phi \rho_i + (1- \rho ) \rho_m$ or
\begin{equation}
\rho (c) = (1-\exp{-n c})\rho_i + \exp{-n c}\rho_m.
\end{equation}
Such mean-field arguments are valid for $c < c_c$ where disconnected clusters of inclusions can be expected to exist. The
size of these patches diverge as $\sim ( c_c - c)^{-\nu }$ and  above the percolation transition macroscopic networks of
patches can be expected to exist which will fundamentally change the material properties of the microalloyed sample.

\section{Discussion}
\label{summary}

We have used the previously obtained understanding of the fundamental plastic instability that leads to the appearance of
shear-bands in amorphous solids to shed light on why and to what extent the addition of
a minute concentration of foreign atoms can defer this instability and improve the toughness of the material.
The mechanism is simple; in the absence of microalloying elements the shear band is formed by a
system spanning line of Eshelby-like quadrupolar displacement fields that combine together to form
a shear-band. To form a displacement field that is a shear-band the quadrupoles must have a uniform and
identical orientation. Isotropic defects interact with the incipient
quadrupoles, forcing them to turn their orientation to minimize the interaction energy. This results in a perturbation of
the
perfect ``in phase" ordering of the quadrupoles that is necessary for the creation of the shear-band. This
perturbation can be overcome by increasing the external shear strain whose effect is to reorganize the quadrupoles to be in
phase. This is precisely the proposed explanation to the data shown in Fig.~\ref{strength}.

The mechanism proposed offers also predictions on the non-spanning nano shear-bands that can be observed
in the strained materials, and their dependence on the concentration of the microalloying elements.
These predictions are easy to check experimentally and we hope that such measurements would be achieved soon.

\section*{Acknowledgement(s)}
This work had been supported in part by and ERC ``ideas" grant STANPAS.
M.M. is grateful to Prof. Raz Kupferman and to Prof. Eran Sharon for their support and useful discussions. M.M. was supported by the Israel-US
Binational Foundation (Grant No. 2010129) and by the Israel Science Foundation (Grant No. 661/13).

\appendix
\section{The numerical protocol}
In our AQS numerical simulations we
use a $50-50$ binary Lennard-Jones mixture to simulate the shear localization discussed in this work. The potential energy for a pair of particles labeled  $i$ and $j$ has the form
\begin{eqnarray}
U_{ij}(r_{ij}) &=& 4\epsilon_{ij}\Big[\Big(\frac{\sigma_{ij}}{r_{ij}}\Big)^{12} - \Big(\frac{\sigma_{ij}}{r_{ij}}\Big)^{6} + A_0 \nonumber\\ &+& A_1\Big(\frac{r_{ij}}{\sigma_{ij}}\Big) + A_2\Big(\frac{r_{ij}}{\sigma_{ij}}\Big)^2\Big] \ , \label{Uij}
\end{eqnarray}
where the parameters $A_0$, $A_1$ and $A_2$ are added to smooth the potential at a scaled cut-off of $r/\sigma = 2.5$ (up to the second derivative). The parameters $\sigma_{AA}$, $\sigma_{BB}$ and $\sigma_{AB}$ were chosen to be $2\sin(\pi/10)$, $2\sin(\pi/5)$ and $1$ respectively and $\epsilon_{AA} = \epsilon_{BB} = 0.5, \epsilon_{AB} = 1$. The particle masses were taken to be equal. The samples were  prepared using high-temperature equilibration followed by a quench to zero temperature ($T=0.001$).
For shearing, the usual athermal-quasistatic shear protocol was followed where each step comprises of an affine shift followed by an non-affine displacement using conjugate gradient minimization. Explicitly, in two dimensions the affine state is achieved by moving the coordinates of every particle $i$ according to
\begin{eqnarray}
&&x_i \to x_i+\delta \gamma y_i\nonumber\\
&& y_i \to y_i \ .
\end{eqnarray}
In an amorphous solid this affine step results in throwing all the forces on the particles out of balance.
Accordingly, to regain mechanical equilibrium, one needs to execute a non-affine step, which is achieved by
gradient energy minimization. Once executed, the next step of straining is done. Note that in such quasi-static protocols
there strain rate is zero. 

The simulations were conducted in two dimensions (2d) and employed Lees-Edwards periodic boundary conditions. Samples were generated  with quench rates ranging from $3.2\times10^{-6}$ to $3.2\times 10^{-2}$ ( in LJ units ), and were strained to greater than $100$ percent. Simulations were performed on system-sizes ranging from $5000$ to $20000$ particles with a fixed density of $\rho = 0.976$ (in LJ units). The simulations reported in the paper have $10000$ particles and a quench-rate of $6.4\times 10^{-6}$ (in LJ units).

\end{document}